\documentstyle[aps]{revtex}
\newcommand{\be}{\begin{equation}}
\newcommand{\ee}{\end{equation}}
\newcommand{\bea}{\begin{eqnarray}}
\newcommand{\eea}{\end{eqnarray}}
\newcommand{\bean}{\begin{eqnarray*}}
\newcommand{\eean}{\end{eqnarray*}}
\begin{document}
\twocolumn[\hsize\textwidth\columnwidth\hsize\csname
@twocolumnfalse\endcsname

\begin{flushright}
ROMA1-2003-1539\\
AEI-2003-063
\end{flushright}

\title{\LARGE\bf Minkowski vacuum \\ in background independent 
quantum gravity}

\author{Florian Conrady${}^{ab}$, Luisa Doplicher${}^a$, Robert
Oeckl${}^b$, Carlo Rovelli${}^{ac}$, Massimo Testa${}^a$
\\[5mm]
\em ${}^a$ Dipartimento di Fisica dell'Universit\`a di Roma ``La
Sapienza", I-00185 Roma.  \\ 
${}^b$ Max-Planck-Institut f\"{u}r Gravitationsphysik,
Albert-Einstein-Institut, D-14476 Golm.\\ 
${}^c$ Centre de Physique Th\'eorique de Luminy, CNRS, F-13288 Marseille. }

\date{\today}

\maketitle
\vspace{1em}

\begin{abstract} 
{We consider a local formalism in quantum field theory, in which no
reference is made to infinitely extended spacial surfaces, infinite
past or infinite future.  This can be obtained in terms of a
functional $W[\varphi,\Sigma]$ of the field $\varphi$ on a closed 3d
surface $\Sigma$ that bounds a finite region $\cal R$ of Minkowski
spacetime.  The dependence of $W[\varphi,\Sigma]$ on $\Sigma$ is
governed by a local covariant generalization of the Schr\"odinger
equation.  Particles' scattering amplitudes that describe experiments
conducted in the finite region $\cal R$ --the laboratory during a
finite time-- can be expressed in terms of $W[\varphi,\Sigma]$.  The
dependence of $W[\varphi,\Sigma]$ on the geometry of $\Sigma$
expresses the dependence of the transition amplitudes on the relative
location of the particle detectors.

In a gravitational theory, background independence implies that
$W[\varphi,\Sigma]$ is independent from $\Sigma$.  However, the
detectors' relative location is still coded in the argument of
$W[\varphi]$, because the geometry of the boundary surface is
determined by the boundary value $\varphi$ of the gravitational field. 
This observation clarifies the physical meaning of the functional
$W[\varphi]$ defined by non perturbative formulations of quantum
gravity, such as the spinfoam formalism.  In particular, it suggests a
way to derive particles' scattering amplitudes from a spinfoam model.

In particular, we discuss the notion of vacuum in a generally
covariant context.  We distinguish the nonperturbative vacuum
$|0_{\Sigma}\rangle$, which codes the dynamics, from the Minkowski
vacuum $|0_{M}\rangle$, which is the state with no particles and is
recovered by taking appropriate large values of the boundary metric. 
We derive a relation between the two vacuum states.  We propose an
explicit expression for computing the Minkowski vacuum from a spinfoam
model.

}

\end{abstract} 
\vskip3pc]

\section{Introduction}

To understand quantum gravity, we have to understand how to formulate
quantum field theory (QFT) in a background-independent manner.  In the
presence of a background, QFT yields scattering amplitudes and cross
sections for asymptotic particle states, and these are compared with
data obtained in the laboratory.  The conventional theoretical
definition of these amplitudes involves infinitely extended spacetime
regions and relies on symmetry pro\-per\-ties of the background.  In a
background independent context this procedure becomes problematic. 
For instance, consider the 2-point function 
\begin{equation}
W(x,y)=\langle 0|\phi(x)\phi(y)|0\rangle.
\label{one}
\end{equation}
In QFT over a background, the independent variables $x$ and $y$ can be
related to the spacetime location of particle detectors.  In a
background independent context, general covariance implies
immediately that $W(x,y)$ is constant for $x\ne y$, and therefore it
is not clear how the formalism can control the localitazion of the
detectors.  (See for instance \cite{wo}.)

Indeed, current efforts to define a quantum theory of gravity
nonperturbatively, such as loop gravity \cite{loop,loopold}, may claim
remarkable theoretical progress, but the problem of deriving
scattering amplitudes remains open.  The effort to develop a covariant
version of loop gravity lead to the spinfoam techniques
\cite{spinfoam}.  These provide well defined expressions for a
Misner-Hawking ``sum over 4-geometries" \cite{Misnerhawking,hawking},
where finiteness results from the discreteness of space revealed by
loop gravity.  The spinfoam formalism provides an amplitude for
quantum states of gravity and matter on a 3d boundary
\cite{hawking,hh}.  But, as far as we know, no formalism is yet
available for deriving particles' scattering amplitudes from these
boundary amplitudes.  Here we indicate a direction to construct such
formalism.  

The key ingredient for developing this formalism is the
Minkowski vacuum state, namely the ``no-particle" state, or the
coherent semiclassical state associated to the classical Minkowski
solution.  The construction of this state is considered a major open
problem in nonperturbative quantum gravity, and it is being studied
using a variety of different techniques.  See for instance
\cite{thomas} and references therein.  Here, we propose a tentative
explicit expression for computing the Minkowski vacuum from a spinfoam
formalism.

We begin by introducing a certain number of general tools, in the
context of the quantum field theory of a simple free massive scalar
field.  The euclidean functional integral over a finite spacetime
region $\cal R$ of spacetime defines a functional $W[\varphi,\Sigma]$
that depends on the field boundary value $\varphi$ and the geometry of
the 3d surface $\Sigma$ that bounds $\cal R$.  We argue that all
physical predictions on measurements performed in the region $\cal R$,
including scattering amplitudes between particles detected in the lab,
can be expressed in terms of $W[\varphi,\Sigma]$.  The geometry of
$\Sigma$ codes the relative spacetime localization of the particle
detectors.  The functional satisfies a local Schr\"odinger equation. 
This defines a covariant formalism for QFT entirely in terms of
boundary data.  For a general formulation of classical and quantum
mechanics along these lines, see \cite{partial,libro} and
\cite{robert}.

Next, we consider the application of this formalism to the
gravitational context.  In the gravitational context, if
$W[\varphi,\Sigma]$ is well defined, then background independence
implies that it is independent from local variations of the location
of $\Sigma$.  At first sight, this seems to leave us in the
characteristic interpretative obscurity of background independent QFT:
the independence of $W[\varphi,\Sigma]$ from $\Sigma$ is equivalent to
the independence of $W(x,y)$ from $x$ and $y$, mentioned above.  But
at a closer look, it is not so: in this context the boundary field
$\varphi$ includes the gravitational field, which is the metric, and
therefore the argument of $W[\varphi,\Sigma] = W[\varphi]$
\emph{still} describes the relative spacetime location of the
detectors!  This fact should allow us to express scattering amplitudes
directly in terms of $W[\varphi]$ even in the background-independent
context.

We distinguish two distinct notions of vacuum.  The first is the
nonperturbative vacuum state $|0_{\Sigma}\rangle$ that the functional
integral on the bulk defines on the (kinematical) Hilbert space
associated to the boundary surface $\Sigma$.  If the metric on
$\Sigma$ is chosen to be to be spacelike, this is the Hartle-Hawking
state \cite{hh}.  In the context we are considering, instead, $\Sigma$
is the boundary of a finite 4d region of spacetime, and
$|0_{\Sigma}\rangle$ is a background-independent way of coding quantum
dynamics \cite{robert}.  The second notion of vacuum is (the local
approximation to) the Minkowski vacuum state.  Here we denote the
Minkowski vacuum state as $|0_{M}\rangle$ (except in equation
(\ref{one}), where it was denoted $|0\rangle$).  We shall argue that
this state is recovered for appropriate values of the boundary metric. 

A main result of this work is an equation connecting the two vacuum
states, and an explicit formula for the Minkowski vacuum state
$|0_{M}\rangle$, in terms of a spinfoam model.  Here we present only
the key ideas and the main results, detailed derivations will appear
elsewhere.

\section{Local tools in QFT}
\subsection{Field to field propagator}
Consider a real massive scalar field $\phi(x)$ on Minkowski space.  To
start with assume it is a free field.  We write $x=(t, \vec x)$. 
Denote by $\varphi(\vec x)$ the classical field configuration at time
zero: $\varphi(\vec x)=\phi(\vec x, 0)$.  The state space at time
zero, ${\cal H}_{t=0}$, is Fock space, where the (distributional)
field operator $\varphi(\vec x)$ and the hamiltonian $H$ are defined. 
The lowest eigenstate of $H$ is the vacuum state $|0_{M}\rangle$, and
its energy $E_{0}$ is zero.  Fock space admits countable bases. 
Choose a basis $|n\rangle$ of eigenstates of $H$ with eigenvalues
$E_n$, and consider the operator
\be %
W(T) = \sum_{n} e^{-TE_{n}}|n\rangle\langle n|.
\label{WT}
\ee %
In the large $T$ limit, this becomes the projector on the vacuum
\be%
\lim_{T\to\infty} W(T) = |0_{M}\rangle\langle 0_{M}|.
\label{vacuum}
\ee%
We now move to a functional Schr\"odinger representation.  Given a
classical field configuration $\varphi$ at time zero, let
$|\varphi\rangle$ be the (generalized) eigenstate of the operator
$\varphi(\vec x)$ with eigenvalue $\varphi$.  We can express any state
$|\Psi\rangle$ of Fock space in this field basis 
\be %
\Psi[\varphi]=
\langle \varphi |\Psi\rangle.  
\ee%
In this representation, the operator
(\ref{WT}) reads 
\be%
W[\varphi_{1},\varphi_{2},T] = \langle \varphi_{1}
|W(T)| \varphi_{2} \rangle.
\ee%
It satisfies the euclidean Schr\"odinger equation (in both variables)
\bea %
-\frac{\partial}{\partial T}\ W[\varphi_{1},\varphi_{2,}T] &=&
H_{\varphi_{1}}\ W[\varphi_{1},\varphi_{2,}T]. 
\eea %
From (\ref{vacuum}), we can obtain the vacuum (up to normalization) as
\be %
 \Psi_{0_{M}}[\varphi] =  \langle\varphi |0_M\rangle = 
 \lim_{T\to\infty} W[\varphi,0,T]. 
\ee %
Particles' scattering amplitudes can be derived from
$W[\varphi_{1},\varphi_{2},T]$.  For instance the 2-point function can
be obtained as the analytic continuation of the Schwinger function
\bea %
S(x_{1},x_{2}) &=& \lim_{T\to\infty} \int D\varphi_1 D\varphi_{2}\ 
W[0,\varphi_{1},T]\  \varphi_{1}(\vec x_{1})\nonumber \\ && 
\hspace{-2em}
W[\varphi_{1},\varphi_{2},(t_{1}-t_{2})]\ \varphi_{2}(\vec x_{2})
\ W[\varphi_{2},0,T]. 
\label{Schwinger}
\eea %
This can be generalized to any $n$-point function where the times
$t_{1}, \ldots t_{n}$ are on the $t=0$ and the $t=T$ surfaces; these
in turn, are sufficient to compute all scattering amplitudes, since
time dependence of asymptotic states is trivial. 

$W[\varphi_{1},\varphi_{2},T]$ admits the well-defined functional
integral representation
\be %
W[\varphi_{1},\varphi_{2},T] = 
\int_{\stackrel
{\scriptstyle{\left.\phi\right|_{\scriptscriptstyle t=T}=\varphi_{1}}}
{\scriptstyle{\left.\phi\right|_{\scriptscriptstyle t=0}=\varphi_{2}}}
}\  \ 
D\phi\ \ e^{-S^{E}_{T}[\phi]}. 
\label{functionalint}
\ee %
Here the integral is over all fields $\phi$ on the strip ${\cal R}$
bounded by the two surfaces $t\!=\!0$ and $t\!=\!T$, with fixed
boundary value.  The action $S^{E}_{T}[\phi]$ is the Euclidean action. 
Notice that using this functional integral representation the
expression (\ref{functionalint}) for the Schwinger function becomes
the well known expression
\be %
S(x_{1},x_{2}) = 
\int D\phi\ \phi(x_{1})\ \phi(x_{2}) \  e^{-S^{E}[\phi]},
\label{wne}
\ee %
obtained by joining at the two boundaries the three functional
integrals in the regions $t\!<\!t_{2}$, $t_{2}\!<\!t\!<\!t_{1}$ and
$t_{1}\!<\!t$.  

The functional $W[\varphi_{1},\varphi_{2},T]$ can be computed
explicitly in the free field theory.  Its expression in terms of the
Fourier transform $\tilde\varphi$ of $\varphi$ is (here
$\omega=\sqrt{\vec k^2 +m^2}$)
\be %
W[\varphi_{1},\varphi_{2},T] = {\cal N}\ 
 e^{{{\textstyle{\scriptscriptstyle{-}\frac{1}{2}}}\!\!\int\!
 {{\frac{d^3k}{(2\pi)^3}}\, \omega\left(
\frac{{|\tilde\varphi_{1}|^2+ 
|\tilde\varphi_{2}|^2}}{\tanh\left(\omega T\right)}
-\frac{{2\tilde\varphi_{1}}
\overline{\tilde\varphi_{2}}}{\sinh\left(\omega T\right)}\right)}}}
\!.\!\!
\label{ffp}
\ee %

\subsection{Kinematical Hilbert space and nonperturbative vacuum}

Consider the 3d surface $\Sigma_{T}=\partial{\cal R}$, namely the
boundary of the strip ${\cal R}$.  This surface is composed by the two
disconnected components $t\!=\!0$ and $t\!=\!T$.  Define a
``kinematical" Hilbert space ${\cal K}_{\Sigma_{T}}$, associated to
the \emph{entire} surface $\Sigma_{T}$, as the tensor product
\be %
{\cal K}_{\Sigma_{T}}= {\cal H}^*_{t=T} \otimes {\cal H}_{t=0}.
\ee %
The notation ${\cal H}^*$, indicates the dual of the Hilbert space
${\cal H}$ (which is of course canonically isomorphic to ${\cal H}$
itself).  Denote as $\varphi=(\varphi_{1},\varphi_{2})$ a field on
$\Sigma_{T}$.  The field basis of the Fock space induces the basis
\be %
|\varphi\rangle =
|\varphi_{1},\varphi_{2}\rangle 
\equiv
\langle\varphi_{1}|_{{}_{t=T}}
\, \otimes\ 
|\varphi_{2}\rangle_{{}_{t=0}}
\ee %
in ${\cal K}_{\Sigma_{T}}$; the vectors $|\Psi\rangle$ of
$H_{\Sigma_{T}}$ are written in this basis as functionals 
\be %
\Psi[\varphi]= \Psi[\varphi_{1},\varphi_{2}] \equiv
\langle\varphi_{1},\varphi_{2}|\Psi\rangle.
\ee %

The functional $W[\varphi_{1},\varphi_{2},T]$ defines the preferred
(bra) state 
\be %
\langle 0_{\Sigma_{T}} |\varphi\rangle\equiv
W[\varphi_{1},\varphi_{2},T].
\label{bra}
\ee %
in this Hilbert space.  This corresponding to the functional $\rho$ of
\cite{robert}.  We call the state $|0_{\Sigma_{T}}\rangle$, the
``nonperturbative vacuum", or ``covariant vacuum".  This state
expresses the dynamics from $t\!=\!0$ to $t\!=\!T$.  A state in the
tensor product of two Hilbert spaces defines a linear mapping between
the two spaces.  The linear mapping from ${\cal H}_{t=0}$ to ${\cal
H}_{t=T}$ defined by $\langle0_{\Sigma_{T}}|$ is precisely the
(imaginary time) evolution $e^{-TH}$.  Indeed, we have by construction
\be %
\langle 0_{\Sigma_{T}}|\ 
\Big(\langle \psi_{out}|\otimes|\psi_{in}\rangle\Big) 
= 
\langle \psi_{out}|e^{-TH}|\psi_{in}\rangle .
\ee %
Or
\be %
\langle 0_{\Sigma_{T}}|\psi_{in}\rangle
= e^{-TH}\ |\psi_{in}\rangle .
\ee %
Notice that the bra/ket mismatch is apparent only, as the three states
live in different Hilbert spaces.

Equation (\ref{vacuum}) shows that in the limit $T\to\infty$ we have 
the projector on the vacuum
\be %
\lim_{T\to\infty} \langle 0_{\Sigma_{T}}|\ 
\Big(\langle \psi_{out}|\otimes|\psi_{in}\rangle\Big) 
= 
\langle \psi_{out}|0_{M}\rangle \ 
\langle 0_{M}|\psi_{in}\rangle .
\ee %
We can therefore write the relation between the two notions of vacuum 
that we have defined as  
\be %
\lim_{T\to\infty}|0_{\Sigma_{T}}\rangle = |0_{M}\rangle\otimes
\langle 0_{M}|.
\ee %
This is a key equation for what follows.  Again, the bra/ket
mismatch is apparent only, as the three states are in different
Hilbert spaces.

The tensor product of two quantum state spaces describes the ensemble
of the measurements described by the two factors.  Therefore ${\cal
K}_{\Sigma_{T}}$ is the space of the possible results of all
measurements performed at time $0$ \emph{and} at time $t$
\cite{partial,libro,robert}.  Observations at two different times are
correlated by the dynamics.  Hence ${\cal K}_{\Sigma_{T}}$ is a
``kinematical" state space, in the sense that it describes more
outcomes than the physically realizable ones.  Dynamics is then a
restriction on the possible outcome of observations
\cite{partial,libro,robert}.  It expresses the fact that measurement
outcomes are correlated.  The state $\langle 0_{\Sigma_{T}}|$, seen as
a linear functional on ${\cal K}_{\Sigma_{T}}$, assigns an amplitude
to any outcome of observations.  This amplitude gives us the
correlation between outcomes at time $0$ and outcomes at time $T$. 
Therefore the theory can be represented as follows.  The Hilbert space
${\cal K}_{\Sigma_{T}}^*$ describes all possible outcomes of
measurements made on $\Sigma_{T}$.  Dynamics is given by the single
linear functional
\begin{eqnarray}
\rho:\hspace{1em} {\cal K}_T &\ \rightarrow \ & C,\nonumber\\ 
|\Psi\rangle &\ \mapsto\ & 
\langle 0_{\Sigma_{T}}|\Psi\rangle .
\end{eqnarray}

For a given collection of measurement outcomes described by a state
$|\Psi\rangle$, the quantity $\langle 0_{\Sigma_{T}}|\Psi\rangle$
gives the correlation probability amplitude between these
measurements.

\subsection{The functional $W[\varphi,\Sigma]$}

We consider the extension this formalism to the case where $\cal R$,
instead of being the strip between two planes, is an arbitrary
\emph{finite} regions of spacetime.  Let $\Sigma$ be the boundary of
$\cal R$, that is a closed, connected 3d surface with the topology
(but in general not the geometry) of a 3-sphere.  Let $\varphi$ be a
scalar field on $\Sigma$ and consider the functional
\be %
W[\varphi,\Sigma] = 
\int_{{\scriptstyle{\left.\phi\right|_{\scriptscriptstyle 
\Sigma}=\varphi}}} D\phi\
e^{-S^{E}_{\cal R}[\phi]}.
\label{functionalintvero} 
\ee %
The integral is over all 4d fields on $\cal R$ that take the value
$\varphi$ on $\Sigma$, and the action in the exponent is the Euclidean
action where the 4d integral is over $\cal R$.  In the free theory the
integral is a well defined Gaussian integral and can be evaluated. 
The classical equations of motion with boundary value $\varphi$ on
$\Sigma$ form an elliptic system and in general has a solution
$\phi_{cl}[\varphi]$, which can be obtained by integration from the
Green function for the shape $\cal R$.  A change of variable in the
integral reduces it to a trivial Gaussian integration times
$e^{-S^{E}_{\cal R}[\varphi]}$.  Here ${S^{E}_{\cal R}[\varphi]}$ is
the field theoretical Hamilton function: the action of the bulk field
determined by the boundary condition $\varphi$.  This function
satisfies a local Hamilton-Jacobi functional equation and solves the
classical field theoretical dynamics \cite{libro,stoyanovski}.

\subsection{Local Schr\"odinger equation}

$W[\varphi,\Sigma]$ satisfies a local functional equation that governs
its dependence on $\Sigma$.  Let $\vec \tau$ be arbitrary coordinates
on $\Sigma$.  Represent the surface and the boundary fields as
$\Sigma: \vec\tau \mapsto x^\mu(\vec\tau)$ and $ \varphi: \vec\tau
\mapsto \varphi(\vec\tau) $.  Let $n^\mu(\vec \tau)$ be the unit
length normal to $\Sigma$.  Then
\be %
n^\mu(\vec \tau)\frac{\delta}{\delta x^\mu(\vec\tau)}W[\varphi,\Sigma]
= H(\vec\tau)\  W[\varphi,\Sigma] 
\label{schroe}
\ee %
where $H(\vec x)$ is an operator obtained by replacing $\pi(\vec x)$
by $-i\frac{\delta}{\delta \varphi(x)}$ in the hamiltonian density
\be %
H(\vec x)= {g}^{-\frac{1}{2}} \pi^2(\vec x)+g^{\frac{1}{2}}\ 
( |\vec\nabla\varphi|^2
+m^2\varphi^2);
\ee %
$g$ is the determinant of the induced metric on $\Sigma$ and the norm
is taken in this metric.  Since $W$ is independent from the
parametrization
\be %
\frac{\partial x^\mu(\vec\tau)}{\partial \vec \tau} \frac{\delta}{\delta
x^\mu(\vec\tau)}W[\varphi,\Sigma] = \vec P(\vec\tau)\ 
W[\varphi,\Sigma]
\ee %
where the linear momentum is $ \vec P(\vec\tau) = \vec\nabla
\phi(\vec\tau)\ {\delta}/{\delta \varphi(\vec\tau)}$.  Details will be
given elsewhere.  If $\Sigma$ is spacelike, (\ref{schroe}) is the
euclidean Tomonaga-Schwinger equation \cite{tomonaga}.  See also the
cautionary remarks in \cite{torre}.

\subsection{Relation with the propagator}

Choose now $\Sigma$ to be a cylinder $\Sigma_{RT}$, with radius $R$
and height $T$, with the two bases on the surfaces $t\!=\!0$ and
$t\!=\!T$.  Given two compact support functions $\varphi_{1}$ and
$\varphi_{2}$, defined on $t\!=\!0$ and $t\!=\!T$ respectively, we can
always choose $R$ large enough for the two compact supports to be
included in the bases of the cylinder.  Then we expect that 
\be %
W[\varphi_{1}, \varphi_{2}, {T}] = \lim_{R\to\infty} \
W[\varphi_{1},\varphi_{2}, \Sigma_{RT}]
\label{Rlimit}
\ee %
because the euclidean Green function decays rapidly and the effect of
having the side of the cylinder at finite distance goes rapidly to
zero as $R$ increases.  Eq (\ref{Schwinger}) illustrates how
scattering amplitudes can be computed from
$W[\varphi_{1},\varphi_{2},T]$.  In turn, eq (\ref{Rlimit}) indicates
how $W[\varphi_{1},\varphi_{2},T]$ can be obtained from
$W[\varphi,\Sigma]$, where $\Sigma$ is the boundary of a finite
region.  Therefore knowledge of $W[\varphi,\Sigma]$ allows us to
compute physical scattering amplitudes.  We expect that this should
remain true in the perturbative expansion of an interacting field
theory as well, where $\cal R$ includes the interaction region.

$W[\varphi,\Sigma]$ can be directly defined in the Minkowski regime as
well.  For a cylindrical box in Minkowski space, let $\varphi=
(\varphi_{out},\varphi_{in},\varphi_{side})$ be the components of the
field on the spacelike bases and timelike side.  Consider the field
theory defined in the box, with time dependent boundary conditions
$\varphi_{side}$, and let $U[\varphi_{side}]$ be the evolution
operator from $t\!=\!0$ to $t\!=\!T$ generated by the (time dependent)
hamiltonian of the theory.  Then $W[\varphi,\Sigma]\equiv\langle
\varphi_{out} |U[\varphi_{side}]| \varphi_{in}\rangle$.  When
$\varphi_{side}$ is constant in time, this can be obtained by analytic
continuation from the Euclidean functional.

\subsection{How far is infinity?}

At first sight, the limits $T,R\to\infty$ seem to indicate that
arbitrarily large surfaces $\Sigma$ are needed to compute vacuum and
scattering amplitudes.  Notice however that the convergence of
$W[\varphi_{1},\varphi_{2},T]$ to the vacuum projector is dictated by
(\ref{WT}): it is exponential in the mass gap $E_{1}$, or the Compton
frequency of the particle.  Thus $T$ at laboratory scales is largely
sufficient to guarantee arbitrarily accurate convergence.  In the
Euclidean, rotational symmetry suggests the same to hold for the $R\to
\infty$ limit.  Thus the limits can be replaced by fixing $R$ and $T$
at laboratory scales.  Problems could arise for the analytical
continuation, which might not commute with the limits, but these
problems do not affect the determination of the vacuum state, where no
analytical continuation is required.

The fact that we can define the vacuum state, or particle states,
locally seems to contradict the fact that the notions of vacuum and
particle states are global.  Let us therefore comment on this delicate
point.  The conventional notions of vacuum and particle states are
global, but particle detectors are \emph{finitely} extended.  In
facts, we may distinguish two distinct notions of particle
\cite{particelle}.  Fock particle states are ``global", while states
detected by a localized detector (eigenstates of local operators
describing detection) are ``local" particles states.  Local particle
states are very close to (in norm), but distinct from, the
corresponding ``global" particle states.  On a flat background, we
conveniently approximate the local particle state detected by the
detectors, using global particle states, which are far easier to deal
with.  The global nature of the conventional definition of vacuum and
particles is therefore an approximation adopted for convenience, it is
not dictated by physical properties of particles we detect.  

Replacing the limits $R\!\to\!\infty$ and $T\!\to\!\infty$ with finite
macroscopic $R$ and $T$ we miss the \emph{exact} global vacuum or
$n$-particle state, but we can nevertheless describe local
experiments.  The restriction of QFT to a finite region of spacetime
must describe completely experiments confined to this region and
states detected by finitely extended particle detectors.

In conclusion, QFT can be formulated in terms of a state space ${\cal
K}_{\Sigma}$ associated to the boundary of a finite region $\cal R$. 
States in ${\cal K}_{\Sigma}$ represent measurement outcomes on
$\Sigma$.  Dynamics is expressed by the single state $\langle
0_{\Sigma}|$ in ${\cal K}_{\Sigma}$, which gives the amplitude for any
complete set of measurements.  This can be computed as a functional
integral over the interior region $\cal R$.  Measurement of the
boundary \emph{field} are represented by the basis $|\varphi\rangle$. 
The Minkowski vacuum state $|0_{M}\rangle$ is obtained by propagation
in imaginary time for a laboratory scale.  \emph{Particle} detection
determines particle states in ${\cal K}_{\Sigma}$, which can be
obtained acting with field operator on $|0_{M}\rangle$.  More details
on boundary particle states will be given elsewhere. 

\section{Quantum gravity}

In quantum gravity, making the formulation described above concrete is
a complex task.  The problem that we consider here is only how to
interpret a functional integral for quantum gravity defining a
functional of the boundary states, assuming this is given to us. 
Concrete definitions of $W[\varphi,\Sigma]$ are rather well developed
in the context of the spinfoam formalism.  Lorentzian and Riemannian
version of the formalism have been studied, and some finiteness
results have been proven to all orders in a perturbative expansion
\cite{finiteness}.

Background independence implies immediately that the gravitational
functional $W[\varphi,\Sigma]$ defined by an appropriate version of
(\ref{functionalintvero}) is independent from any local variation of
$\Sigma$.  Fixing the topology of $\Sigma$, we have therefore
\be %
W[\varphi,\Sigma]=W[\varphi].
\ee %
At first sight, this seems the sort of independence from position and
time, that renders background-independent QFT difficult to interpret. 
The independence of $W[\varphi,\Sigma]$ is indeed analogous to the
independence of $W(x,y)$ from $x$ and $y$ mentioned at the beginning
of this paper.

However, the property of $\Sigma$ that codes the relative spacetime
location of the detectors is the \emph{metric} of $\Sigma$.  In the
gravitational case, the metric of $\Sigma$ is not coded in the
location of $\Sigma$ on a manifold: it is coded in the boundary value
of the gravitational field on $\Sigma$.  Therefore the relative
location of the detectors, lost with $\Sigma$ because of general
covariance, comes back with $\varphi$, as this includes the boundary
value of the gravitational field.  Therefore, if we are given a
functional integral for gravity, we can interpret it exactly as we did
for the scalar field!  The boundary value of the gravitational field
plays the double role previously played by $\varphi$ and $\Sigma$.  In
fact, this is precisely the core of the conceptual novelty of general
relativity: there is no a priori distinction between localization
measurements and measurements of dynamical variables.

$W[\varphi]$ determines a preferred state $|0_{\Sigma}\rangle$,
defined by $ \langle 0_{\Sigma}|\varphi \rangle = W[\varphi]$ in the
kinematical state space $\cal K$ associated with the boundary.  This
is the covariant vacuum, and codes the dynamics.  It satisfies a
dynamical equation analogous to equation (\ref{schroe}), where
$H(\vec\tau)$ is now the hamiltonian constraint density operator.  But
since $W$ is independent from $\Sigma$ by general covariance, the left
hand side of (\ref{schroe}) vanishes, leaving
\be %
H(\vec\tau)\ W[\varphi]=0,
\ee %
which is the (lorentzian) Wheeler-DeWitt equation \cite{hawking}.

\subsection{Minkowski vacuum in quantum gravity}

The quantum state $|0_{M}\rangle$ that describes the Minkowski vacuum
is not singled out by the dynamics alone in quantum gravity.  Rather,
it is singled out as the lowest eigenstate of an energy $H_{T}$ which
is the variable canonically conjugate to a nonlocal function $T$ of
the gravitational field defined as the proper time along a given
worldline.  

This situation has an analogy in the simple quantum system formed by a
single a relativistic particle.  In the Hilbert space of such a system
there is no preferred vacuum state.  But we can choose a preferred
Lorentz frame, and therefore a preferred Lorentz time $x^0$.  The
conjugate variable to $x^0$ is the momentum $p_{0}$, and there is a
(generalized) state of minimum $p_{0}$.

To find the Minkowski vacuum state, we can repeat the very same
procedure used above.  The only difference is that the bulk functional
integral is not over the bulk matter fields, but also over the bulk
metric.  This difference has no bearing on the above formulas, which
regard the boundary metric, which, in the two cases, is an independent
variable.

As a first example, a boundary metric can be defined as follows. 
Consider a three-sphere formed by two ``polar" $in$ and $out$ regions
and one ``equatorial" $side$ region.  Let the matter+gravity field on
the three-sphere be split as 
\be %
\varphi=(\varphi_{out},
\varphi_{in},\varphi_{side}).
\ee%
Fix the equatorial field $\varphi_{side}$ to take the special value
$\varphi_{RT}$ defined as follows.  Consider a cylindrical surface
$\Sigma_{RT}$ of radius $R$ and height $T$ in $R^4$, as defined above. 
Let $\Sigma_{in}$ (and $\Sigma_{out}$) be a (3d) disk located within
the lower (and upper) basis of $\Sigma_{RT}$, and let $\Sigma_{side}$
the part of $\Sigma_{RT}$ outside those disks, so that 
\be %
\Sigma_{RT}= \Sigma_{in}\cup \Sigma_{out} \cup \Sigma_{side}.
\ee%
Let $g_{RT}$ be the metric of $\Sigma_{side}$ and let
$\varphi_{RT}=(g_{RT},0)$ be the boundary field on $\Sigma_{side}$
determined by the metric being $g_{RT}$ and all other fields being
zero.  Given arbitrary values $\varphi_{out}$ and $\varphi_{in}$ of
all the fields, included the metric, in the two disks, consider
$W[(\varphi_{out},\varphi_{in},\varphi_{RT})]$.  In writing the
boundary field as composed by three parts as
$\varphi(\varphi_{out},\varphi_{in},\varphi_{side})$ we are in fact
splitting $\cal K$ as 
\be ä 
{\cal K} = H_{out}\otimes H^*_{in} \otimes
H_{side}.
\ee ä
Fixing $\varphi_{side}=\varphi_{RT}$ means contracting the
covariant vacuum state $|0_{\Sigma}\rangle$ in $\cal K$ with the bra
state $\langle \varphi_{RT}|$ in $H_{side}$.  For large enough $R$ and
$T$, we expect the resulting state in $H_{out}\otimes H^*_{in}$ to
reduce to the Minkowski vacuum.  That is 
\be%
\lim_{R,T\to\infty} \langle \varphi_{RT}|0_{\Sigma}\rangle=
|0_{M}\rangle \otimes \langle 0_{M}|. 
\ee%
Therefore for a generic $in$ configuration, and up to normalization
\be%
\Psi_{M}[\varphi] =
\lim_{R,T\to\infty} W[(\varphi,\varphi_{in},\varphi_{RT})].
\label{vuotogr}
\ee%
gives the vacuum functional for large $R$ and $T$.  (Below we shall
use a simpler geometry for the boundary.)  

One may hope that the convergence in $R$ and $T$ is fast.  These
formulas allow us to extract the Minkowski vacuum state from a
euclidean gravitational functional integral.  $n$-particle scattering
states can then be obtained by generalizations of the flat space
formalism, and, if it is well defined, by analytic continuation in the
single variable $T$.  Notice that we are precisely in the case of time
independent $\varphi_{side}$, where analytical continuation may be
well defined.

\subsection{Spinnetworks and spinfoams}

The argument of $W$ is not a classical field: it is an element of the
eigenbasis of the field operator.  In the gravitational case,
(functions of) the gravitational field operator can be diagonalized,
but eigenvalues are not continuous fields.  In loop quantum gravity,
eigenstates of the metric are spin network states $|s\rangle$. 
Therefore the quantum gravitational $W$ must be a function of spin
network states $W[s]$ on $\Sigma$, and not of continuous gravitational
fields on $\Sigma$.  In fact, this is precisely what a spin foam model
provides.

A spinfoam sum where the degrees of freedom are not cut off by the
choice of a fixed triangulation is defined by the Feynman expansion of
the QFT over a group, studied in \cite{groupfieldtheory}. 
Let us recall here the basic equations of the formulations, referring
to \cite{groupfieldtheory} and \cite{libro} for motivations and
details.  Let $\Phi(g_{1},\ldots,g_{4})$ be a field on $[SO(4)]^4$,
satisfying
\begin{equation} \label{ginv}
	\phi(g_1,g_2,g_3,g_4) = \phi(g_1g,g_2g,g_3g,g_4g), 
\end{equation}
for all $g \in {\rm SO}(4)$.  Consider the action
\begin{eqnarray}
S[\phi] &=& \frac{1}{2} \int (\phi)^{2} \ + 
\frac{\lambda}{5!} \int
(P_{{}_H}\phi)^5. 
\end{eqnarray}
Here $P_{{}_H}$ is defined by 
\begin{eqnarray} 
\label{Hoper}
	P_{{}_H}\phi(g_1,g_2,g_3,g_4) &=&\nonumber \\
&&	\hspace{-6em}
	\int_{H^4} dh_{1}\ldots dh_{4} \
	\phi(g_1h_{1},g_2h_{2},g_3h_{3},g_4h_{4}).
\end{eqnarray}
where $H$ is a fixed $SO(3)$ subgroup of $SO(4)$, and 
$\int \phi^5$ is a short hand notation for 
\begin{eqnarray}
\int \phi^5 &=& \int \prod_{i=1}^{10} dg_i
~~\phi(g_1,g_2,g_3,g_4)\phi(g_4,g_5,g_6,g_7)\nonumber \\
&&\hspace{-3em}
 \ \phi(g_7,g_3,g_8,g_9)
\phi(g_9,g_6,g_2,g_{10}) \ \phi(g_{10},g_8,g_5,g_1).
\label{actionGQFT}
\end{eqnarray}
The Feynman expansion of this theory is a sum over spinfoams and can
be interpreted as a well-defined version of the Misner-Hawking sum
over geometries.  Transition amplitudes between quantum states of
space can be computed as expectation values of $SO(4)$ invariant
operators in the group field theory.  In particular, the boundary
amplitude of a 4-valent spin network $s$ can be computed as
\be%
W[s]=\int D\Phi\ f_{s}[\Phi] \ e^{-S[\phi]},
\ee%
The spinfoam polynomial is defined as
\begin{eqnarray} 
f_{s}[\phi] &=& \prod_{n} 
\int dg_{n_{1}}\ldots dg_{n_{4}}\
R^{(j_{n_{1}})}_{\alpha_{n_{1}}}{}^{\beta_{n_{1}}}(g_{n_{1}}) \ldots 
\nonumber \\
&&
R^{(j_{n_{4}})}_{\alpha_{n_{1}}}{}^{\beta_{n_{4}}}(g_{n_{4}}) \ 
v^{i_{n}}_{\beta_{n_{1}}\ldots\beta_{n_{4}}}
\ \prod_{l}\ \delta^{l_{1}l_{2}}
\end{eqnarray} 
where $n_{1},\ldots,n_{4}$ indicate four links adjacent to the node
$n$, and $n_{i}=l_{1}$ (or $n_{i}=l_{2}$) if the $i$-th link of the
node $n$ is the outgoing (or ingoing) link $l$.

We can now implement equation (\ref{vuotogr}) in this theory.  Instead
of the cylindrical boundary consider above, we can choose a simpler
geometry.  Let the spin network $s'$ be composed by two parts
connected to each other, $s'=s\#s_{T}$.  Let $s$ be arbitrary and
$s_{T}$ to be is a weave state \cite{weave} for the three-metric
$g_{T}$ defined as follows.  Take a 3-sphere of radius $T$ in $R^4$. 
Remove a spherical 3-ball of unit radius.  $g_{T}$ is the three-metric
of the three-dimensional surface (with boundary) formed by the sphere
with removed ball.  I recall that a weave state for a metric $g$ is an
eigenstate of (functions of the smeared) metric operator, whose
eigenvalues approximate (functions of the smeared) $g$ at distances
large compared to the Planck length.

The quantity
\be%
\Psi_{M}[s]= \langle s |0_{M}\rangle = \lim_{T\to\infty}\int D\Phi\
f_{s\#s_{T}}[\Phi] \ e^{-S[\Phi]}.
\label{formula}
\ee%
is then a tentative ansatz for the quantum state describing the
Minkowski vacuum in a ball of unit radius.  This quantity can be
computed explicitly \cite{groupfieldtheory} and may be finite at all
orders in $\lambda$ \cite{finiteness}.

\section{Conclusions}

In this paper we have sketched several general ideas on the physical
interpretation of the formalism in background independent QFT. The
main ideas we have considered are the following

\begin{itemize}
    \item[(i)] In QFT, the functional integral over a finite region
    defines the functional $W[\varphi,\Sigma]$ of the boundary field,
    which expresses the physical content of the theory.

    \item[(ii)] This functional can be used to compute the vacuum
    state $|0_{M}\rangle$, taking choosing $\Sigma$ appropriately.

    \item[(iii)] In a background independent theory, $n$ particle
    functions $W(x_{1},\ldots, x_{n})$ become meaningless, because
    they are independent from the coordinates; while
    $W[\varphi,\Sigma]$ maintains its physical meaning, in spite of
    the fact that it is independent from $\Sigma$.  This is because in
    a gravitational theory the relative location of the detectors is
    coded in $\varphi$ and not in $\Sigma$.  Localization measurements
    are on the same footing as the dynamical variables measurements.
   
\item[(iv)] The functional $W[\varphi]$ defines a state
    $|0_{\Sigma}\rangle$ that codes the dynamics of the theory by
    determining the correlation amplitudes between boundary
    measurements.

\item[(v)] The Minkowski vacuum state $|0_{M}\rangle$ can be computed
from nonperturbative quantum gravity by choosing appropriate boundary
values of the gravi\-tational field.
    
\item[(vi)] A tentative formula giving the Minkowski vacuum state in
terms of a spinfoam model is given by equation (\ref{formula}).
    
\item[(vii)]  Relevant analytical continuation is in the proper length of
    the boundary, not in the time coordinate.

\end{itemize}  

Much remains to be done and many points are far from clear.  The most
urgent of these problems is the following.  The spinfoam model we have
referred to in the text is Riemannian, not Euclidean.  Namely its
amplitudes correspond to the complex quantity $e^{iS_{E}}$, where
${S_{E}}$ is the Euclidean action, and not to a real exponential.  The
relation between the Euclidean, Riemannian and Lorentzian spinfoam
models has not yet completely clear, we believe.

\vspace{2em}
\centerline{--------}
Thanks to Daniele Oriti for clarifications on spinfoams.  FC thanks
the Daimler-Benz Foundation and DAAD for support.  CR and FC thank the
Physics Department of the University of Roma for hospitality.

\end{document}